\documentstyle[12pt,aaspp4,tighten]{article}
\begin{document}

\begin{center}
\LARGE \bf {AN UPDATED THEORETICAL SCENARIO FOR GLOBULAR CLUSTER STARS}
\end{center}
\begin{center}
\vspace{0.5cm}
S. Cassisi$^{1,2}$, V. Castellani$^{3,4}$ S. Degl'Innocenti$^{3,5}$ and
A. Weiss $^{6}$
\end{center}
\vspace{0.4cm}
{\it $^{1}$ Osservatorio Astronomico di Collurania, via M.Maggini, 
64100 Teramo, Italy}\\[0.2cm]
{\it $^{2}$ Dipartimento di Fisica, Universit\'a de L'Aquila, Via 
Vetoio, 67010 L'Aquila, Italy}\\[0.2cm]
{\it $^{3}$ Dipartimento di Fisica, Universit\'a di Pisa, 
Piazza Torricelli 2, 56100 Pisa, Italy}\\[0.2cm]
{\it $^{4}$ Istituto Nazionale di Fisica Nucleare, LNGS, 67010 Assergi, L'Aquila}\\[0.2cm]
{\it $^{5}$ INFN
Sezione di Ferrara, via Paradiso 12, 44100 Ferrara, Italy}\\[0.2cm]
{\it $^{6}$ Max Plank Institut for Astrophysics, Karl Schwarzschild
Strasse 1, D-85470 Garching b. Munchen, Germany}\\[0.2cm]

\begin{center}
Received ............   Accepted..............
\end{center}
\vspace{2.5 cm}
Cassisi et al.: An updated evolutionary scenario for GC stars\\[1.cm]
{\it Send offprints requests to: V. Castellani, 
 Dipartimento di Fisica, Universit\'a di Pisa, 
Piazza Torricelli 2, 56100 Pisa, Italy }\\[0.4cm]
Main Journal:\\[0.4cm]
Thesaurus code numbers: 08.05.3, 08.06.3, 08.07.1, 08.08.2\\[0.4cm]
Proofs to:  V. Castellani, Dipartimento di Fisica, Universit\'a di Pisa, 
Piazza Torricelli 2, 56100 Pisa, Italy
\newpage
\begin{abstract}

In the first part of this paper we revisit the history 
of theoretical predictions
for HB luminosities in  old Population II stellar clusters, starting from
the results of `old' evolutionary computations to introduce in 
various steps all the available `new' physics. We discuss the influence 
of physical ingredients on selected evolutionary parameters,  finally 
presenting  models which incorporate  
all the most recent updating of the relevant physics. 
The evolutionary  behavior of such models is extensively
investigated for selected choices about the cluster metallicity, 
discussing theoretical predictions concerning both cluster isochrones and the
calibration of the parameter R in terms of the original amount of
He in stellar matter.  One finds that the `new' physics has a relevant
influence on  both these parameters, moving in particular cluster ages
to a much better agreement with current cosmological evaluations.
This scenario is implemented by a further set of stellar models where 
element diffusion is taken into account. The comparison between 
theoretical scenarios with or without diffusion is presented and
discussed.  A discussion of current observational 
constraints to the light of the updated theory closes the paper.

\end{abstract}

{\bf keywords:} Stars:evolution - Stars:general - Stars:fundamental parameters
- Stars:horizontal-branch
\section{Introduction}

Since galaxies were  born in an already expanding Universe, the age
of the Universe appears as a safe upper limit for the age of
any star and any stellar cluster. The fact that several 
determinations of globular cluster ages yielded values larger than the age
of the Universe as based on current evaluations of the Hubble constant
(see, e.g., Van den Bergh 1994, Tanvir et al.\ 1995) has stimulated  
a renewed interest in the theory of globular cluster Pop.~II stars. 
At the same time, significant improvements in the input physics needed
for stellar evolution have been made, such that noticeable changes of
the theoretical results could be expected. These improvements initially
were motivated by the results of helioseismology, which opened a new
window into the interior of the Sun, allowing for an extremely accurate
determination of the solar structure. This provided a severe challenge
for stellar structure theory. The efforts undertaken resulted in a new
generation of opacity data (Rogers \& Iglesias 1992; Seaton et al.\
1994, Iglesias \& Rogers 1996) and equations of state (Mihalas et al.\ 1990; Rogers et al.\
1996), which led to a much better prediction of solar oscillations
and also resolved many long-standing problems in our understanding of
pulsating stars. In addition, helioseismology required particle
diffusion to be taken into account in solar models (see Bahcall et
al.\ 1995 and references therein).

Stimulated by the success with regard to the Sun, the new opacities
and equation of state, along with improvements in low-temperature
opacities (e.g.\ Alexander \& Fergusson 1994), nuclear
cross-sections and neutrino emission rates, have been applied now to
low-mass metal-poor stars in order to investigate the above-mentioned
age problem.
Several investigations (Chaboyer \& Kim 1995;  Mazzitelli et al. 1995: MDC; 
VandenBerg et al. 1996;
D'Antona et al. 1997; Salaris et al. 1997: Paper I) have already shown that
models with some of the updated physics inputs tend to predict lower
cluster ages, thus at least decreasing the size of the discrepancy,
if not resolving it.  
The new physics still needs to be applied to more massive and more
metal-rich stars, although some of it, e.g.\ opacities, already are in
use (Bono et al. 1997 a, 1997b, 1997c) However, the full 
consequences of all improvements have not yet
been evaluated. As an example we mention the evolution and pulsations
of Cepheid stars.

In the present paper we are concerned with Pop.~II stars only.
We intend to follow a twofold purpose. Firstly,
we will present stellar models appropriate for globular cluster
studies that include {\it all} of the improvements listed above. These
models will cover the complete relevant mass and metallicity range,
and include all evolutionary stages from the zero-age main sequence
until the end of the helium-burning phase on the horizontal
branch. Our calculations therefore provide the most up-to-date set of
stellar models applicable to all problems of globular cluster dating.
In particular, we show for the first time how particle diffusion
influences the evolution of low-mass stars until the end of the
horizontal-branch phase. In our feeling, in the future
diffusion will become an
integral part of modern stellar evolution, such as it already has been
for several years in the case of solar models.

Secondly, we will demonstrate how each of the various steps in
improving the input physics is influencing the models. This is important
because of the variety of calculations available in the literature
 that include {\it some}
but not all of the new physics. In order to compare these results, it
is necessary to be able to translate the differences in physical
assumptions into differences in stellar properties.
In the first part of this paper we will approach this problem,
starting from a suitable set of ``old'' evolutionary computations 
and introducing, step by step, the available ``new'' physics in order
to make  
clear the influence of the new assumptions on selected evolutionary 
parameters. At the end of Sect.~2, we will finally present our best models 
which will incorporate  all the most recent improvements
in the relevant physics. However, these models will still be
calculated ignoring element diffusion.

In Sect.~3 evolutionary predictions for these best models will then be
extensively 
investigated for selected choices about the cluster metallicity, presenting
theoretical predictions concerning cluster isochrones. This will be
repeated in Sect.~4 for  
a set of stellar models where element diffusion is
properly taken into account. The comparison between theoretical
scenarios with and without diffusion will be presented and
discussed.  Section 5 will finally deal with a discussion of the
influence on the R-parameter and the consequences for the inferred
original amount of helium in stellar matter. We will also critically
discuss the theoretical uncertainties on R. Finally, the conclusions
follow.

\section{ Input physics  and Pop.II models }

As a starting point we will  assume as a {\it reference} frame (step  1) 
the evolutionary scenario presented by Straniero \& Chieffi (1991) and by
Castellani et al. (1991: hereinafter CCP), 
which covers with a homogeneous
set of computations the major evolutionary phases experienced by 
galactic globular cluster stars. As a relevant point, let us here recall that
the above evolutionary scenario appears in excellent agreement with 
computations based on similar physics  given by Sweigart (1987);
in particular theoretical predictions concerning the mass of the He core at
the He ignition agree within few thousandths of solar mass.

The `step 1' column in Table 1 gives
details on the relevant physics adopted in those models and which
has been submitted to successive modifications. Top to bottom one finds:\\
- Equation of State (EOS) Str88: Straniero (1988) implemented at the
 lower temperature with Saha equation.\\
- Radiative Opacity for H, He mixtures (OPAC). LAOL: Los Alamos Opacity 
Library (Huebner et al. 1977) implemented at the lower 
temperature with Cox \& Tabor (1976) opacity tables.\\
- Radiative Opacity for C, O mixtures (OPAC-CO). LAOL: Los Alamos Opacity 
Library (Huebner et al. 1977). \\
- He burning rates ($\alpha$-rates). Fow75: Fowler et al. (1975), 
    Harris et al. (1983), Caughlan et al. (1985).\\
- Neutrino energy losses (NEU). Mun85: Beaudet et al.(1967),
 Munakata et al. (1985), Richardson et al. (1982).

Electron screening (Graboske et al. 1973, DeWitt et al. (1973) 
and electron conductivity (Itoh et al. 1983) have not been submitted
to relevant improvements since that time. As a matter of fact,
numerical experiments performed with our code show that 
neither improvements in strong electron screening, as given by 
Itoh et al. (1977) and Itoh et al. (1979), nor
the alternative  approach to weak and intermediate screening (Mitler 1977)
do affect the evolutionary phases we are dealing with. 

Table 1 gives a list of the various steps performed in the input physics
together with the corresponding values for selected evolutionary
quantities. The upper portion of the table gives the steps in
updating the physics inputs, whereas
in the  lower portion of Table 1 one finds  selected results
concerning the H burning phase of a 0.8$M_{\odot}$ model (Y=0.23,
Z=0.0001) and the He burning phase  of the same model but assuming
the original mass reduced to 0.7 $M_{\odot}$ by mass loss.  Top to bottom 
one finds: the luminosity (LogL$^{TO}$) and the age (t$^{TO}$) of the 
0.8$M_{\odot}$  H burning model at the track Turn Off (TO), 
the luminosity (LogL$^{flash}$), the age (t$^{flash}$) and the mass (M$_c$)
of the He core at the He flash and the surface helium abundance 
(Y$_{HB}$) after the first dredge-up. For the He burning  0.7 $M_{\odot}$ 
one finally finds the  Zero Age Horizontal Branch luminosity (LogL$_{ZAHB}$),
and effective temperature (LogTe$_{ZAHB}$) together with the time t$_{HB}$ 
spent in the central He burning phase as a Horizontal Branch (HB) star.
Here, as well as through all this paper, luminosities and masses are in
solar units.

As shown in Table 1, the updating of the input physics runs as
follows: 

i) EOS from Str88 to OPAL (Rogers 1994, Rogers et al. 1996),
 implemented in the temperature-density region
not covered by OPAL with Str88, plus Saha EOS in the outer stellar
layers. The transition from OPAL to other EOS appears smooth and
without discontinuities. 

ii) OPAC and OPAC-CO from LAOL to OPAL (Rogers \&
Iglesias 1992, Iglesias \& Rogers 1996), 

iii) $\alpha$-rates from Fow75 to Cau88 (Caughlan \& Fowler 1988) and, 
finally, 

iv) NEU from Mun85 to Haft94 (Haft et al. 1994) for plasma neutrino
production and Itoh et al. (1996) 
for the other kinds of neutrino energy losses.

 Even a quick inspection of results in Table 1 shows the
relevant effects produced by the OPAL-EOS on the MS lifetimes and
TO-luminosities, an occurrence already well discussed in the 
literature (see, e.g., Chaboyer \& Kim 1995). For HB models,  
one finds that
improvements in the opacity of H-rich mixtures have the major effect
of moderately increasing the HB luminosity ($\Delta$LogL$\sim$0.02)
and decreasing the HB lifetime by 3.4 \% .  As expected, CO opacity
affects only the advanced phases of central He burning, decreasing the
HB lifetimes by a further 7\% . As a whole, one finds that the major
effect of the new opacities is the decrease of HB lifetimes by the not
negligible amount of about 10 \%. Step 4 in Table 1 shows that the
passing from the previous EOS to the more recent OPAL EOS does not
affect HB lifetimes; however one finds that the HB luminosity
increases by a further $\Delta$LogL$\sim$0.02, in spite of the of the
small decreases in Mc, whereas the age of the flashing RG decreases by
about 2 Gyr. 

Steps 5, 6, 7 and 8 finally report the effect of improved evaluations of
the triple $\alpha$ nuclear reactions and of the plasma neutrino 
energy loss rates. On very
general ground, one expects that both these mechanisms affects the He
ignition at the flash, affecting in turn the structure of the
initial ZAHB models. To disentangle this effect from the effect 
on the physics of HB models, step 5 and 6 concern only ZAHB models, 
introducing the new rates for 3$\alpha$ reactions (Caughlan
\& Fowler 1988) and for plasma neutrino production (Haft et al. 1994)
in two subsequent steps {\it for the fixed value of the ZAHB Helium
core mass} given by the result of step 4.  One finds that the new
3$\alpha$ rates further increase, though slightly, the HB luminosity,
 whereas HB lifetimes are again substantially decreased by a further 8\%.
 On the contrary, one finds that HB structures are only marginally
affected by the NEU treatment, as early predicted (Gross 1973).

Step 7 shows the effect of new 3$\alpha$ rates on H burning models
as HB progenitors. Step 8  gives finally the results for our
`best' models where all the available updating of the physics have been 
taken into account. Due to the effect of both 3$\alpha$ rates and
NEU, the He flash is delayed and the top  luminosity of the RG 
structures is increased, becoming about 0.2 mag brighter than
in Straniero \& Chieffi (1991; step 1 in Table 1). Correspondingly
the value of Mc `jumps' from 0.5054  M$_{\odot}$ to  0.5152 M$_{\odot}$,
contributing to a further increase of the HB luminosity. 
From data in Table 1, one recognizes that 3$\alpha$ rates and NEU give a
similar contribution to the quoted increase of Mc. As a whole, one finds 
that passing from CCP to present best models the major modifications
about HB evolution appear an increase of the ZAHB luminosity
by about $\Delta$LogL$\sim$0.06 ($\sim$0.15 mag.) whereas HB lifetimes are
decreased by the huge amount of, about, 23 \%. As one can easily
understand, and as we will discuss later on, this will have rather 
dramatic  effects on current calibration of the 
R parameter. 

To orientate the reader in the current literature, let us review
available theoretical estimates in terms of the quoted physical
scenarios.  As a starting point, let us notice that CCP computations 
adopt more or less improved input physics than adopted in
previous computations (as, e.g., Sweigart 1987, Dorman \& VandenBerg
1989, Lee \& Demarque 1990).  Dorman (1992) adopts  neutrino
energy losses and
opacities as in CCP, improving nuclear reactions rates as in Caughlan
\& Fowler (1988) but taking the EOS from Eggleton et al.  (1973).
 Dorman et al. (1993) adopt the same inputs as 
Dorman (1992), but low-temperature opacities from Alexander (1975, 1981).
Mazzitelli et al. (1995) have OPAL EOS
and opacity, but using Dappen et al. (1988) EOS in H burning models 
(as stated in D'Antona et al. 1997 who updated the turn off models
with OPAL EOS); nuclear reactions rates are from Harris
et al. (1993) and neutrinos from Itoh et al. (1989).  Salaris et al. (1997)
 models overlap present step 4
assumptions. As a result, one finds that our step 8 is till present
the only one including all available updating of the input
physics. According to such an evidence, in the following section we
will investigate the evolutionary behavior of similar models,
discussing the calibration of the most relevant evolutionary
parameters.

\section {`Best' canonical models }

The evolutionary behavior of our `best' models, as defined by step 8 in
the previous section, has been investigated for selected
choices on the assumed star metallicity and adopting everywhere
an original amount of He given by Y=0.23 as a suitable value for 
population II stars. In all cases we assumed solar scaled composition
as given by Grevesse \& Noels (1993). However, alpha-enhanced distributions 
can be taken into account bearing in mind the scaling law discussed
by Salaris et al. (1993). It is worth noting that the 
validity of such relation has been recently questioned by VandenBerg \&
Irwin (1997), but for a metal-rich regime ([Fe/H]$>$ -0.8) and for large 
$\alpha$-enhancement factors ([$\alpha$/Fe]$>$ 0.3; see also Weiss, Peletier
\& Matteucci, 1993, for the same topic), i.e., for values beyond the range
suitable for globular cluster stars.

Table 2 gives selected data
 of the models at the track turn-off (TO) for the various choices on the
stellar mass and for the metallicity Z=0.0001, 0.0002,
0.001 and 0.006. Left to right one finds: the metallicity (Z) , the mass of the model
(M), the age (t$^{TO}$), the luminosity (LogL$^{TO}$) and the effective temperature
 (LogTe$^{TO}$) at the track Turn Off (TO).

On the basis of these evolutionary tracks 
H burning isochrones have been computed for the quoted assumed 
metallicities and covering the range of ages suitable
for galactic globular cluster stars. 
Table 3 gives detailed informations on the isochrone TO luminosity and effective
temperature.  Left to right one finds:
 the metallicity (Z),  the age (t$^{TO}$),
 the luminosity (LogL$^{TO}$), the effective temperature
 (LogTe$^{TO}$) and the mass of the model (M$^{TO}$) 
at the isochrone Turn Off (TO). As  expected, data for the case Z=0.0002 overlap
similar computations presented in  Paper I, since passing from
step 4 to step 8 affects only the advanced evolution of 
RG and HB structures. Thus present
computations may be regarded as an extension to larger 
metallicities of the quoted computations.

 We agree with the comment of our unknown referee about the risk
of using TO luminosity as a parameter to derive cluster ages. 
From an observational point of view it appears quite difficult
to define this parameter with high accuracy (Richer, Fahlman \&
VandenBerg 1988); the average uncertainty on the TO magnitude
can be estimated of the order of $\simeq$ $\pm$ 0.10 mag, which leads 
to an indetermination on the derived age of the order of $\pm$1.5
Gyr (see Chaboyer, Demarque \& Sarajedini 1996 for a discussion
on this point). Moreover, from a theoretical point of view, 
being the TO defined as the bluest point on the isochrone, 
the determination of the TO magnitude Mv(TO) is related
to the isochrone color (Chaboyer 1995, Chaboyer et al. 1996),
thus depending on the stellar effective temperature, i.e., on the stellar
radius, which can be affected by significant uncertainties,
depending on the theoretical treatment of convection in
superadiabatic layers (MDC). 
For such reasons, the
use of other age indicators (see, e.g., Chaboyer et al. 1996) has 
been suggested in several investigations. However, since such a parameter
is still widely in use, let us discuss with some details our results
on the matter, at least to allow a comparison with previous results
appeared in the literature. 

The best fit of the data for the dependence 
of the TO luminosity on the cluster ages 
gives
the analytical relations:\\

Logt$_9$ = -0.946 LogL$^{TO}$ + 1.465  (Z=0.0002).\\

Logt$_9$ = -1.117 LogL$^{TO}$ + 1.414  (Z=0.001). \\

Logt$_9$ = -1.239 LogL$^{TO}$ + 1.325  (Z=0.006).\\

\noindent where t$_9$ is the cluster age in Gyr. These relations  reproduce 
the computational results in the range 8 to 18 Gyr
with a maximum error of few 10$^8$ years.
As a result, one finds that, for each given age, the TO luminosities are predicted
to decrease with metallicity  with a slope in fair agreement with a large body 
of previous predictions but with lower predicted luminosities
for each given age. This is shown in Fig. 1, which 
compares the dependence of present TO luminosities 
on the assumed metallicity for a given age (t=12 Gyr) with
similar results  already appeared in the literature. 
In the figure (as well as in some other following figures) theoretical
expectations for O-enhanced  mixtures (Bergbusch \& VandenBerg 
1992, VandenBerg 1992, Dorman et al. 1993) are reported versus the
total fraction of heavy elements, as obtained by the same quoted papers.

One finds that
present results represent in all cases a lower boundary for
current evaluations of TO luminosities, thus decreasing current age 
estimates for a given TO luminosity. More in details, 
present results  predict TO luminosities systematically  lower by 
about $\Delta$LogL$\simeq$0.05  with respect to previous computations, 
but Mazzitelli, D'Antona \& Caloi (1995: MDC) who, 
independently of the adopted treatment of superadiabatic
convections, predict again larger luminosities but with a difference
from present results which decreases as the metallicity increases.
According to the previous analytical relations, one easily finds 
that   the quoted decrease $\Delta$LogL$\simeq$0.05 implies a corresponding 
decrease by about 10\% in previous estimates of globular cluster ages.
We will further discuss this point in the final section.

Beyond the problem of TO luminosities, H burning models deserve 
further attention as progenitors of He burning models, determining
the structural parameters which will constrain the evolutionary behavior and,
in particular, the luminosity of HB stars.
Computational results concerning those parameters are reported in Table
4 for the labeled assumptions about the cluster metallicity. 
Left to right one finds: the metallicity (Z), the mass (M$_c$) of the
He core at the He flash, the surface helium abundance
(Y$_{HB}$) after the first dredge-up, the age (t$^{flash}$) and the luminosity
(LogL$^{flash}$) at the He flash and the mean value between the minimum and
the maximum in luminosity (LogL$^{bump}$) during the RGB `bump'. 
Data in Table 4 will allow the approach of He burning phases adopting
self-consistent evolutionary values for the two parameters
characterizing a ZAHB structure, namely the mass of the He core
(Mc) and the He abundance in the stellar envelope. Since both 
values depend only marginally on the assumptions made about the cluster ages,
Table 4 reports the values corresponding to a 0.8 M$_{\odot}$ evolving Red Giant
which can be safely assumed as representative of theoretical
expectations in a sufficiently large range of ages. 
Here let us notice that the discussed increase (see Table 1) of 
the predicted luminosity 
of the RG tip would affect current estimate of the Hubble 
constant H$_0$ when  using such a feature as a distance indicator. As a 
matter of the fact, one easily finds that the quoted increase 
by 0.2 mag in the top RG luminosity implies an increase by about
10\% in the distance and, in turn, a decrease by the same amount
of the H$_0$ estimate.

The amount of extra He ($\Delta$Y) brought to the stellar 
surface by the first dredge up appears in good agreement with similar
evaluations already given in the literature (see, e.g., 
Castellani \& Degl'Innocenti 1995 and references therein).
 Figure 2 compares present 
masses of the He cores in the flashing Red Giants with previous
results on that matter. Again one finds that all current
evaluations, but MDC, have a rather similar dependence on
the assumed metallicity. However, one finds that our `best'
models  predict Mc values in all cases larger than previous predictions, 
an occurrence which acts in the sense of increasing the expected
luminosity of ZAHB structures.

By adopting M$_c$ and $\Delta$Y values from H burning models we are now
in the position of predicting the evolutionary behavior of He burning
Horizontal Branch (HB) structures. 
Table 5 gives detailed informations on the HR diagram location of Zero Age 
Horizontal Branch (ZAHB) together with a comparison between present and
CCP He burning lifetimes. 
Left to right one finds: the mass (M), the luminosity
(LogL) and the effective temperature (LogTe) of the zero-age horizontal-branch,
ZAHB, model (following CCP we assume as ZAHB structures the models already evolved
by 1 Myr), the time (t$_{He}$) spent during the central He burning (until the disappearance
of the convective core) and the same quantity (t$_{He}^{CCP}$) but for the CCP
models.

As expected on the basis of the exploratory
computations given in the first part of this paper, one finds that the
expected luminosity of ZAHB models is substantially increased whereas 
He burning lifetimes in all cases fall down by more than 20\%. Figure 3
presents predictions about the luminosity of the ZAHB model in the RR Lyrae
instability strip (LogTe=3.85) with previous results on that matter. 
One finds that
`old' computations, as given by CCP or Lee \& Demarque (1990) agree in
predicting lower luminosities, by about $\Delta$LogL$\approx$0.05. 
This occurrence  implies that, when using ZAHB models as `standard candles' to constrain
 the cluster distance modulus (DM), `old'
 computations would produce smaller DM, thus lower luminosities of the
observed TO and, finally, larger ages. The same figure shows that
all the more recent computations agree in predicting more luminous
ZAHBs. In particular one finds that at the lowest metallicity, we predict
luminosities in close agreement with MDC, notwithstanding the (small)
difference in the He core masses. Note that the difference at 
the larger metallicities
can be understood in terms of the different slope of the
 M$_c$-metallicity relation already disclosed in Fig. 2.

Figure 4 finally compares present He-burning lifetimes with the ones given
in CCP and with the value originally predicted by Buzzoni et al. (1983)
for the two assumed metallicities. The emerging scenario concerning
current evaluation of the amount of original He in globular cluster
stars will be discussed in Sect. 5. 

\section{Element diffusion}

A general discussion on the relevance of element diffusion in
the evolution of Pop.II stars has been already given in a 
previous paper (Castellani et al. 1997: Paper II) and it will be not 
repeated here. Here we only recall that the diffusion of both
He and heavy elements is taken into account, according to the 
algorithm adopted in Thoul et al. (1994). Table 6 gives selected 
evolutionary properties of models for the various choices on the
stellar mass and for the investigated metallicities. 
As a whole, our results closely follows the general trends discussed by Proffitt \&
VandenBerg (1991) in their pioneering paper to which we address the
reader for a general discussion. 
Figure 5 compares the run in the HR diagram of isochrones with and without
element diffusion for a selected metallicity and for the labeled choices
on the cluster age
whereas Table 7 gives detailed informations on the isochrone TO luminosity and effective
temperature.
Tables 6 and 7 are the homologous of Tables 2 and 3 
 given in the previous section for computations neglecting
element diffusion. Figure 6 shows the dependence of isochrones TO
luminosity on cluster age, as compared with similar results but
without allowing for element sedimentation.

In the case of sedimentation the best fitting of the data
 connecting  ages to TO luminosities gives
for ages between 8 and 18 Gyr the relations:\\

Logt$_9$ = -0.924 LogL$^{TO}$ + 1.414   (Z=0.0002)\\

Logt$_9$ = -1.130 LogL$^{TO}$ + 1.384   (Z=0.001) \\

Logt$_9$ = -1.170 LogL$^{TO}$ + 1.288   (Z=0.006) \\

\noindent which again reproduce the results within a few 10$^8$ years.

According to the discussion given in Paper II
for the case Z=0.0004, one finds that element diffusion moderately decreases
theoretical expectations for TO luminosities for each given cluster age.
Figure 6 now shows that such an effect depends on the assumed metallicity,
increasing when the metallicity is decreased. As a matter of the fact, 
the effect of diffusion on the TO luminosity is larger in the most
metal poor isochrones since the diffusion in the stellar envelopes is
larger due to the thinner convective envelopes on MS stars.
At the lowest metallicity
(Z=0.0001) $\Delta$LogL$\approx$0.04: thus, for a given observed TO luminosity,
allowing for sedimentation would decrease age estimates by about 10\%.
On the contrary, when Z=0.006 one expects rather negligible variations.

Table 8 gives selected structural parameters of models at
the He flash, to be used as inputs to the He-burning models (for a discussion
 of the effects of diffusion on the luminosity of the RGB `bump', see also
Cassisi et al. (1997).
On this basis we present in Fig. 7 a comparison between 
HB evolution with or without diffusion;
in the same figure the path in the HR diagram of these models is also compared
with similar results from CCP.
Table 9 gives 
details on the ZAHB structures and on the corresponding helium
 burning evolutionary times.
One has to advice that HB structures presented in both Fig. 7 and Table 9
assume a 0.8 M$_{\odot}$ model as H-burning progenitor. However, Table 8
shows that increasing, e.g., the RG masses (thus decreasing the cluster age)
M$_c$ decreases but Y$_{HB}$ increases, with balancing effects 
on the predicted HB luminosities. As a result, numerical experiments
disclosed that HB data based on a 0.8 M$_{\odot}$ progenitor can be
safely taken as representative of HB models in the range 0.7 $<$ M/M$_{\odot}$
$<$ 0.9, at least, thus covering quite a large range of cluster ages.

Figure 8 summarizes the results of this paper relevant for the problem
of cluster ages, disclosing the run with metallicities of ZAHB and TO
luminosities, with or without diffusion, and for selected choices
about the assumed ages. In this figure, data concerning the TO luminosity
have been implemented with similar data but for Z=0.0004, as given 
in Paper II with a `step 4' physics which is fully compatible
with present computation as far central H burning models are concerned.
As already discussed, present HB are brighter than estimated in 
Paper II (see Table 1).
However, the same Fig. 8 shows that present computations
keep predicting that diffusion decreases the HB luminosity by
about $\Delta$LogL$\sim$0.02, in agreement with the results
of Paper II.

Figure 9 shows the calibration of age in terms of
the difference in luminosities between ZAHB (taken at LogTe=3.85) 
and TO, as predicted with or
without diffusion, and as compared with original predictions in
CCP. As discussed in Paper II one finds that diffusion plays a minor
role in that calibration. However, the same Fig. 9 shows that the
new physics, as a whole, reduces by about 4 Gyr theoretical
calibrations based on the old physics.

The detailed comparison between theory and observation is a delicate
question, beyond the purpose of this theoretical paper. However, 
one may test present predictions vis-a-vis recent estimates 
of HB luminosities derived by recent Hipparcos parallaxes measurements. 
This is shown in Fig. 10, which compares the data presented
by De Boer et al. (1997) with our theoretical predictions 
converted in Mv, B-V magnitudes according to model atmospheres
by Kurucz (1992). One finds that the HB luminosity level appears
in quite good agreement with the quoted observations. Here we only notice
that the two stars which lie below the ZAHB around B-V$\simeq$ 0 
both have been  corrected for a rather large reddening (E(B-V)=0.10),
one lacking -to our knowledge- of recent metallicity estimates.

As for the outcome of the improved theoretical scenario, let us 
recall that in Paper I it has been already shown that in the 
step 4 scenario the color magnitude diagram of a typical metal poor 
galactic globular can be reproduced by a 12 Gyr (no diffusion) isochrone. 
Taking into account
that, at LogTe=3.85, our best HB models without diffusion turn out
 to be more luminous by $\Delta$LogL$_{ZAHB}\approx$ 0.02, one estimates
that the `new' age shifts toward 11 Gyr without sedimentation, and even 
 below if sedimentation is taken into account.

Figure 11 discloses that present results foresee a rather low dependence
of $\Delta$LogL(HB-TO) on the cluster metallicity. As a matter of
the fact, assuming, e.g. an age of 12 Gyr one finds that passing from
Z=0.0002 ([Fe/H]=-1.97) to Z=0.001 ([Fe/H]=-1.27) we predict an
increase in $\Delta$LogL(HB-TO) corresponding to $\approx$0.08 mag,
 independently
of any assumptions about the efficiency of sedimentation, this difference
increasing when the cluster age is decreased.

\section {The parameter R}

\noindent
Since the pioneering paper by Iben (1968) one knows that
evolutionary predictions on the evolution of Pop.II stars can be used
to constrain the amount of original He in globular cluster stars.
Calibrations of the R parameter, i.e., 
the number ratio between HB stars and RG
more luminous than the {\it HB luminosity level} have been given
by Buzzoni et al. (1983) and, more recently, by Caputo, Martinez Roger 
\& Paez (1987)
and by Bono et al. (1995). According to current estimates, observational 
values for R appear ranging around R ${\sim}1.1$. In terms of the
quoted calibrations this implies $Y{\sim}0.23$, which is consequently the
value currently adopted in discussing globular cluster stars.

\noindent However, the evolutionary results discussed in the previous sections
deeply affects such a scenario. We already found that the updated
physics moderately affected (increased) theoretical expectations
about HB luminosities, largely decreasing  HB lifetimes.
According to such an evidence, one expects a decreasing value
of R and -thus- a larger  predicted value of Y for any given value
of R.   Owing to the relevance of the argument, let us  derive 
a quantitative evaluation of R as given by updated  predictions
about evolutionary times both along the RG and through the HB
evolutionary phases. It has been already found that
 evolutionary times along the upper 
portion of the RG branch show a rather negligible dependence on both the
chemical composition (within Pop.II limits) and mass of the
evolving stars (see e.g. Castellani \& Castellani 1993, Bono et al. 1995,
 Salaris \& Cassisi 1997).
 Now we find a small but not negligible dependence on
the efficiency of sedimentation. By best fitting  computational
results we find in the interval 1.5$\leq$LogL$\leq$1.8:\\

t$^{flash}$-t = 730.93 -629.14 LogL +144.73 (LogL)$^2$ (No Diffusion) \\

t$^{flash}$-t = 732.93 -625.73 LogL +143.32 (LogL)$^2$ (Diffusion) \\

\noindent where t$^{flash}$-t represents the time (in 10$^6$ yr) spent by a RG 
above the luminosity L.
However, when Z=0.006 these relations can be 
safely used only in clusters with age lower than, about, 13 Gyr.
At larger ages, the clump of stars along the RG branch becomes 
fainter than the HB luminosity level, as disclosed by data in the previous
Table 8, and the relations would require a correction to properly
account for such an occurrence (see Bono et 
al. 1995 for a discussion on that matter).

According to the procedure envisaged by Bono et al. (1995) we will take
as reference  the luminosity  level of the ZAHB at LogTe=3.83,
evaluating the time spent by RG stars above such a luminosity and
taking HB evolutionary lifetimes from the 
models starting HB evolution at that effective temperature.
Table 10 gives data about these two ingredients together with the
corresponding estimates of R for the labeled choices on the
metallicity, with or without allowing for the efficiency of
sedimentation.  Top to bottom one finds: the luminosity
(LogL$_{ZAHB}$) of the ZAHB model at LogTe=3.83 the time (t$_{He}$)
spent by the same model during the central He burning (until the
disappearance of the convective core), the time (t$_{RG}$) spent by 
RGB stars above LogL$_{ZAHB}$, the value (R(3.83)) of the corresponding
R parameter and the same values but when the ZAHB luminosity level
is artificially increased by $\Delta$LogL=0.05 (R(3.83)+0.05) 
and 0.1 (R(3.83)+0.1). As already known,  one recognizes that an increase
of the metallicity tends to slightly increase the expectations on R for a
given value of Y. Focusing, e.g., our attention on the case
Z=0.001, one finds that when Y=0.23 the theoretical prediction
given by Bono et al. (1995), R= 1.19, should now be decreased to
R= 1.05 for model without sedimentation or to R= 0.95 if sedimentation is
taken into account. According to all available calibrations of R one finds 
$\delta$Y$\approx$0.4 $\delta$LogR . As a consequence, 
present evolutionary scenario would predict that current estimate of 
original He should be increased
by about $\Delta$Y$\sim$0.02 if sedimentation is neglected, or
by about $\Delta$Y$\sim$0.04 with sedimentation at work. 
As a result, observational data already interpreted in the
literature as an evidence for Y= 0.23 should now lead to
the rather unpalatable conclusion Y$\simeq$ 0.27. 

However, before entering in a discussion of the values in Table 10, one has
to notice that the calibration  of R depends on He-burning evolutionary times
which, in turns, are mainly governed by the poorly determined
cross section for $^{12}$C + $\alpha$ reaction (see also Dorman 1992). 
Along this paper we 
adopted for He burning reactions the rates given by Caughlan 
\& Fowler (1988) which should improve previous evaluations given by
the same authors in 1985. Comparison between these two rates shows a 
rather negligible differences in the triple alpha rates, but a large
decrease in the $^{12}$C + $\alpha$ rates which, in turn, largely
contributes  to
the decrease of HB lifetimes one finds in Table
1 between steps 4 and 7. As a matter of fact, 
 about 60\% of this decrease in HB
lifetime (and of the corresponding decrease in the predicted value
of R) can be attributed to these new rates.
However, errors estimates on such a cross section 
are still as large as a factor of two, containing in this range also previous
 estimates given by Caughlan et al. (1985). Moreover,
numerical experiments performed on HB models adopting recent reaction rates
by Buchmann (1996), with errors estimates still of about 70\%, tends
to move the lifetimes toward the values estimated in old computations,
based on Caughlan et al. (1985).
One can only conclude that theoretical calibrations of R in terms of
Y are still affected by too large errors to be used for accurate
calibrations of such a relevant parameter, and that the values
of R given in Table 10 are still affected by theoretical errors corresponding
to an error on Y of about $\Delta$Y$\sim$0.02-0.03. If one adds the 
further errors related to the observational procedure, i.e., the errors
on the HB luminosity level, on the bolometric correction for the
corresponding  RG stars and on the star counts (see, e.g., Brocato et al.
1995) one should conclude that R still appears as a too risky parameter
to allow evaluations of Y with a reasonable accuracy.

Last  two rows in Table 10 finally give theoretical evaluations for R 
when the adopted luminosity level is artificially increased above the
ZAHB level by $\Delta$LogL=0.05 and 0.1, respectively. These 
values can be used to evaluate theoretical expectations  on R
when the mean luminosity of RR Lyrae is taken instead of the ZAHB
luminosity as reference luminosity level. In the meantime these values
give an estimate of the error on Y produced by observational
errors in that level.  One easily finds that an overestimate
by $\Delta$LogL=0.05 (0.125 mag.) will produce an overestimate of
He by about $\Delta$Y$\sim$0.015.
Note that previous evaluations of R appear  only as a lower
limit for theoretical expectations  for cluster 
with blue  HB. Less massive,
hot HB structures have He burning evolutionary times increased
by 20\% or more (see Fig. 7 and Castellani et al. 1994), 
with a corresponding increase on theoretical 
expectations about the parameter R.

\section{Conclusions.} 

In this paper we have followed the evolution of theoretical
predictions concerning Pop.II stellar models vis-a-vis the recent
progresses in the input physics. Stellar models including all
the more recent evaluations of theoretical ingredients have been
presented and discussed, with particular regard to the problem of
globular cluster ages. We found that similar models tend to decrease
previous estimates about the cluster age.  The account for element
sedimentation goes in the same direction.  As a whole one finds that
`canonical age estimates', as given in CCP, have to be decreased by
about 4 Gyr, promising a much better agreement with cosmological
constraints.  
We finally drove the attention on the large indetermination of the
theoretical procedure adopted to constrain the cluster original abundance
of He, concluding that accurate results on that matter must wait for
a better determination of the nuclear cross section  $^{12}$C + $\alpha$.

Detailed tabulations on both evolutionary tracks and/or cluster
isochrones are available upon request by E-mail.\\

{\bf Acknowledgments}

It is a pleasure to thank F. D'Antona for  her comments on a preliminary 
draft of the paper.


\pagebreak 


\newpage

\section{Figure caption}

\figcaption [] {Behavior of the TO luminosity on the assumed metallicity for a
given age (t=12 Gyr). Results for present `best'  canonical models are
compared with similar results available in the literature. For the MDC 1995
models CM indicates the adoption by the authors of the Canuto \& Mazzitelli (1991)
treatment of overadiabatic convection while MLT indicates the adoption of the
usual mixing lenght theory. 
}
 
\figcaption [] {He core masses at the He flash as a function of metallicity 
for present models (canonical and with element diffusion) as compared with
 similar data already appeared in the literature.
}
 
\figcaption [] {The ZAHB luminosity  at $\log{T_e}=3.85$, as a function of
metallicity for  present models, compared with previous results,
as labeled.}
 
\figcaption [] {Central He-burning lifetimes as a function of the ZAHB
effective temperature for present
models (solid line) compared with similar data in CCP (dashed line)
and with the  predictions by Buzzoni et al. (1983) for
HB models with $\log{T_e}=3.83$ (stars). Metallicities as labeled.}
 
\figcaption [] {H burning isochrones for Y=0.23, Z=0.001
and for the labeled ages for the present models
without and with element diffusion (upper and lower panel, respectively). 
The time interval between consecutive isochrones is 1 Gyr.
Note that the standard isochrones are calculated until a luminosity
lower than that of the helium flash.}
 
\figcaption [] {Dependence of the TO luminosity on the cluster ages for
the three labeled metallicities. The results for canonical models
(dashed line) are compared with similar results but for models
with element sedimentations (solid line).}
 
\figcaption [] {Comparison of the HB evolution for models with
(solid line) and without (dashed line) diffusion for Z=0.001 and Y=0.23.
Similar results from CCP (dot-dashed line) are also shown.}
 
\figcaption [] {TO luminosities for selected labeled ages and  ZAHB luminosities 
at $\log{T_e}=3.85$ as a function of metallicity, for models with (solid line)
and without (dashed line) element diffusion.}
 
\figcaption [] {The calibration of age in terms of the difference in
luminosities ($\Delta\log{L}\rm (HB-TO)$) between ZAHB (at $\log{T_e}=3.85$) and TO, 
as predicted by present models with (solid line) and without diffusion
(dashed line) and Z=0.0002. The results are compared with original
predictions by CCP (dotted line).}
 
\figcaption [] {{\it a}) Theoretical ZAHB for standard models compared with Hipparcos
estimates of HB magnitudes from De Boer et al. 1997 (see text). When available,
labeled metallicities are from Gray et al. 1996. {\it b)}As {\it a}) but for models computed
 by accounting for element diffusion.}
 
\figcaption [] {The dependence on metallicity of the difference in luminosities
between ZAHB (at LogTe=3.85) and TO ($\Delta$LogL(HB-TO)),
 as predicted by  present
models with (solid line) or without diffusion (dashed line) for selected
labelled ages.  Present results are compared with original predictions in
CCP (dotted line).}

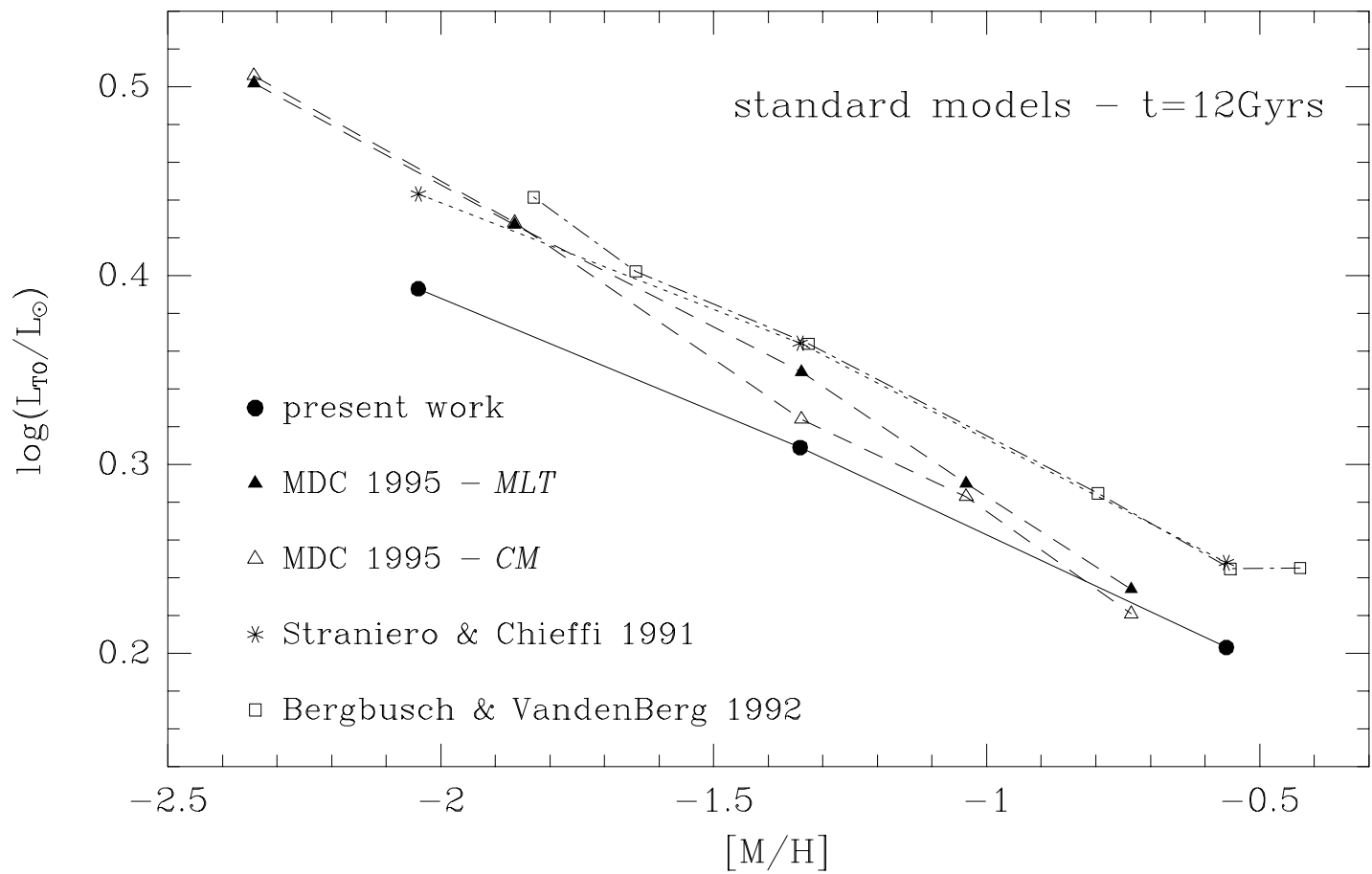
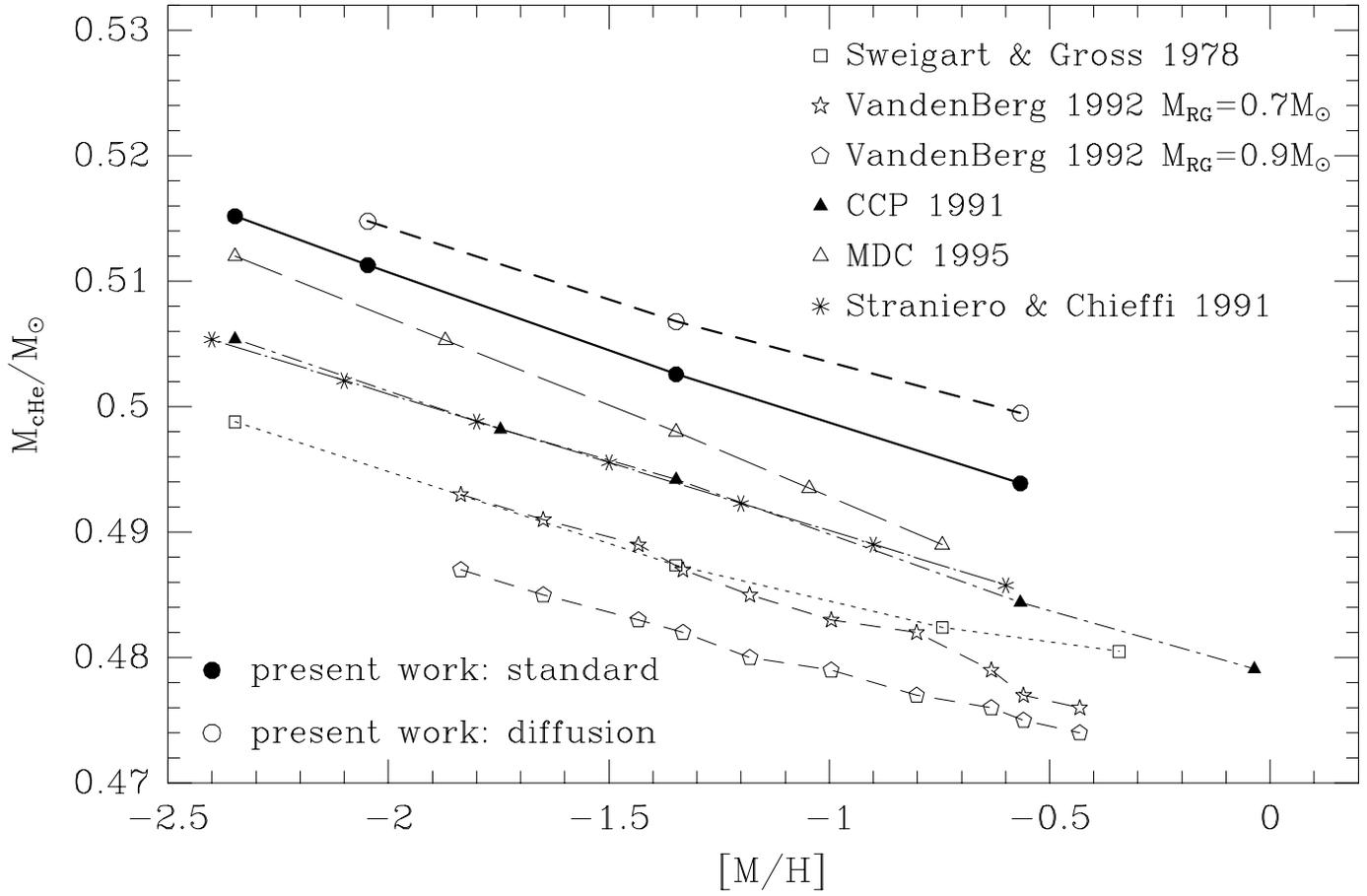
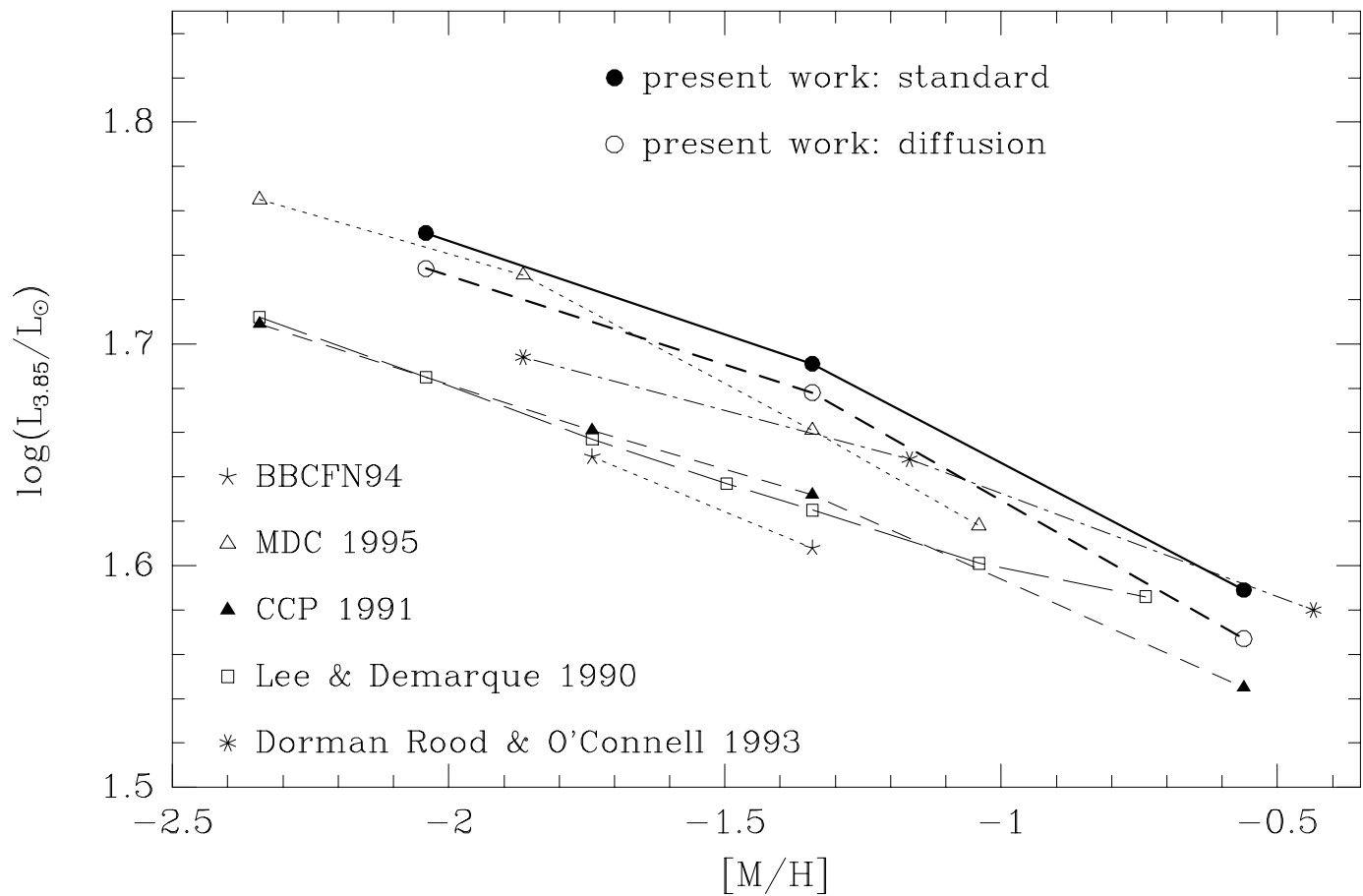
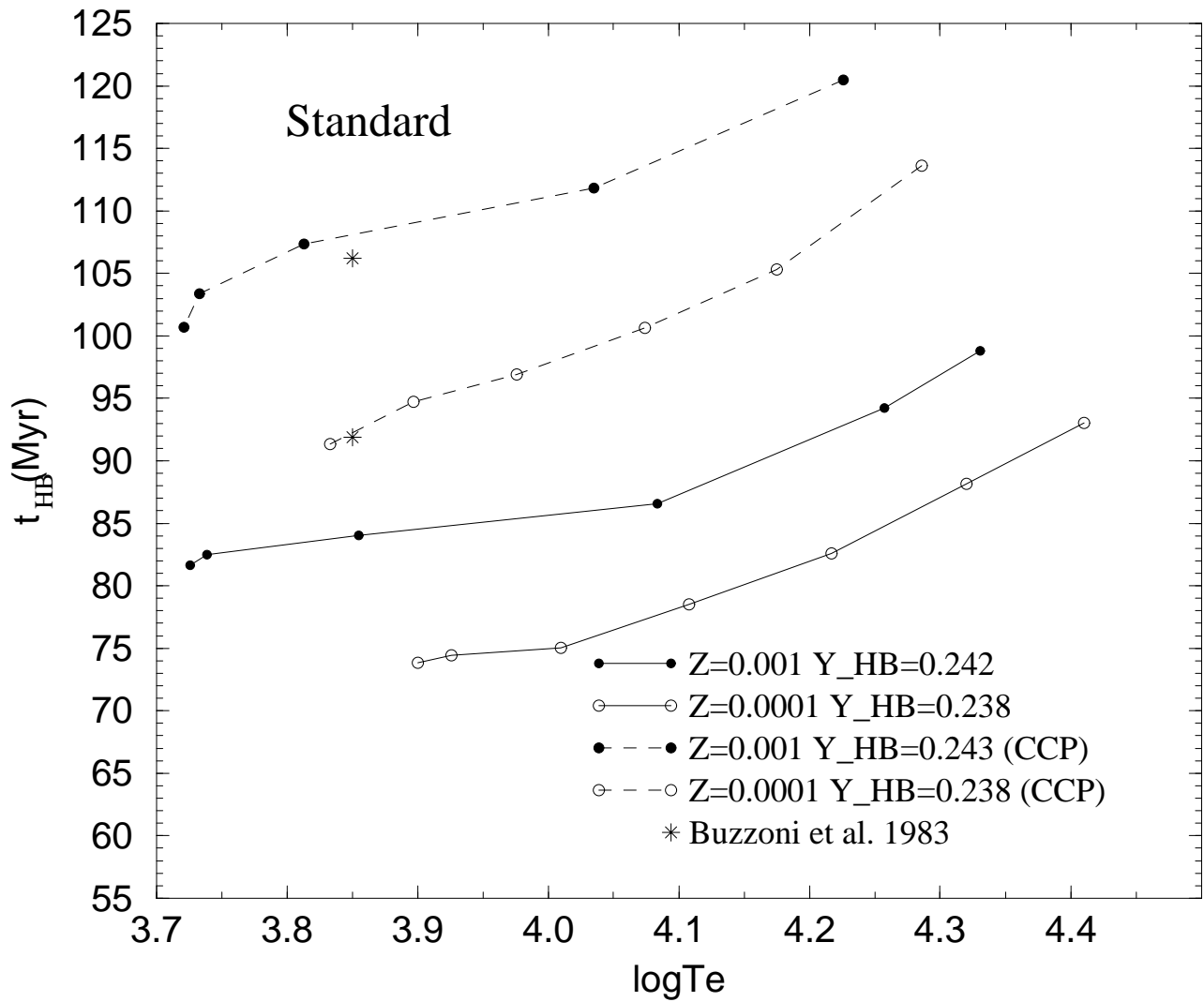
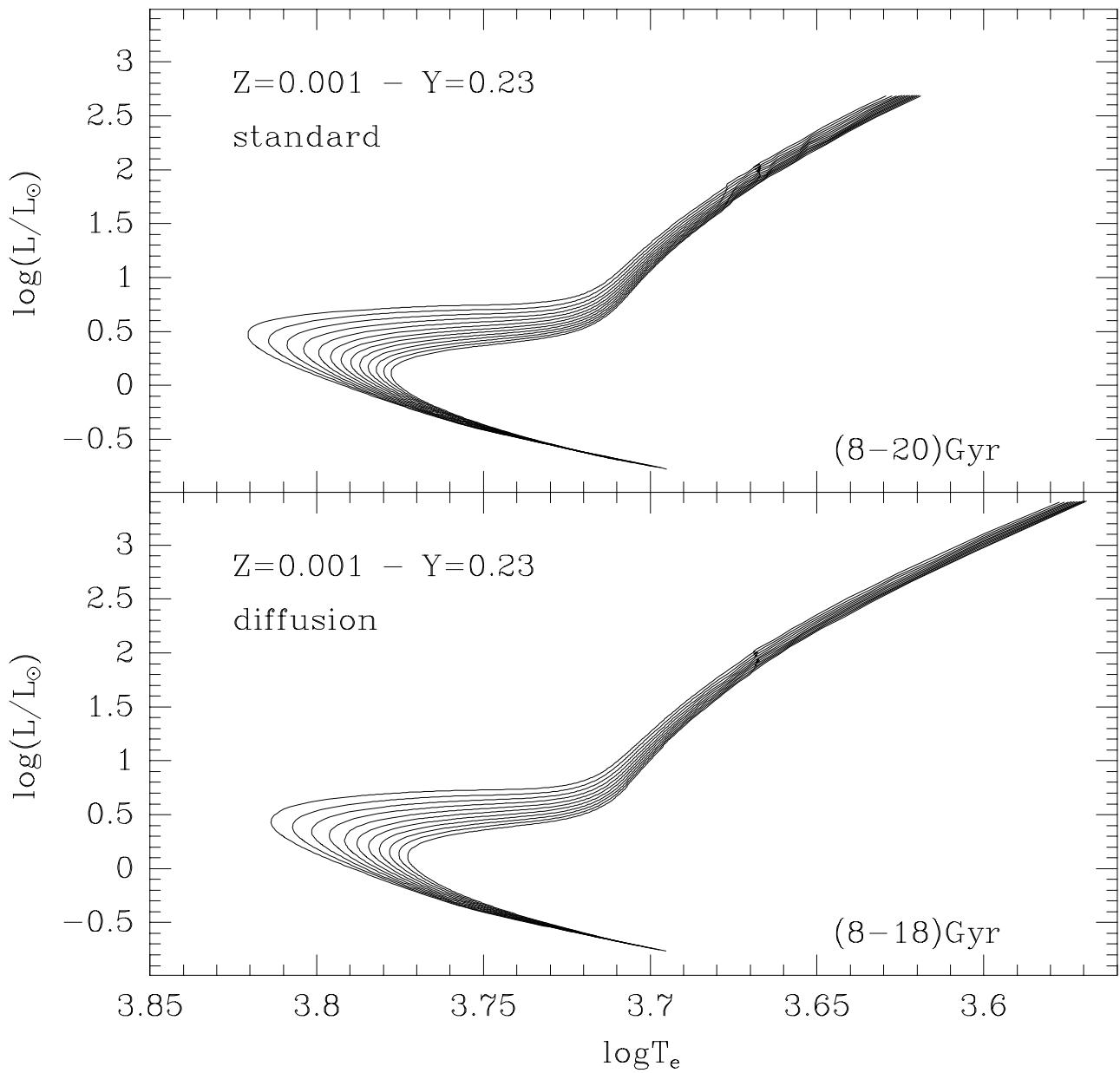
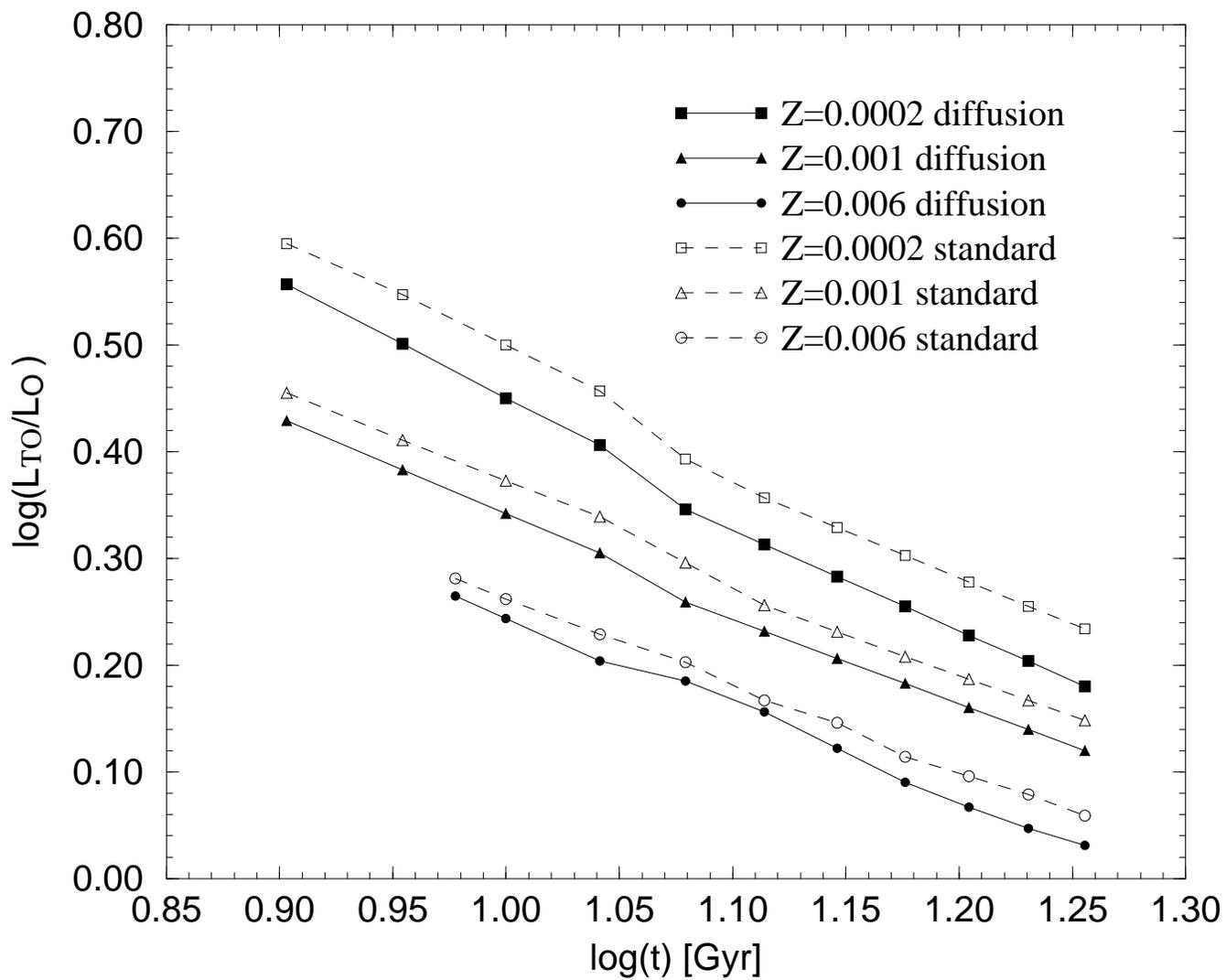
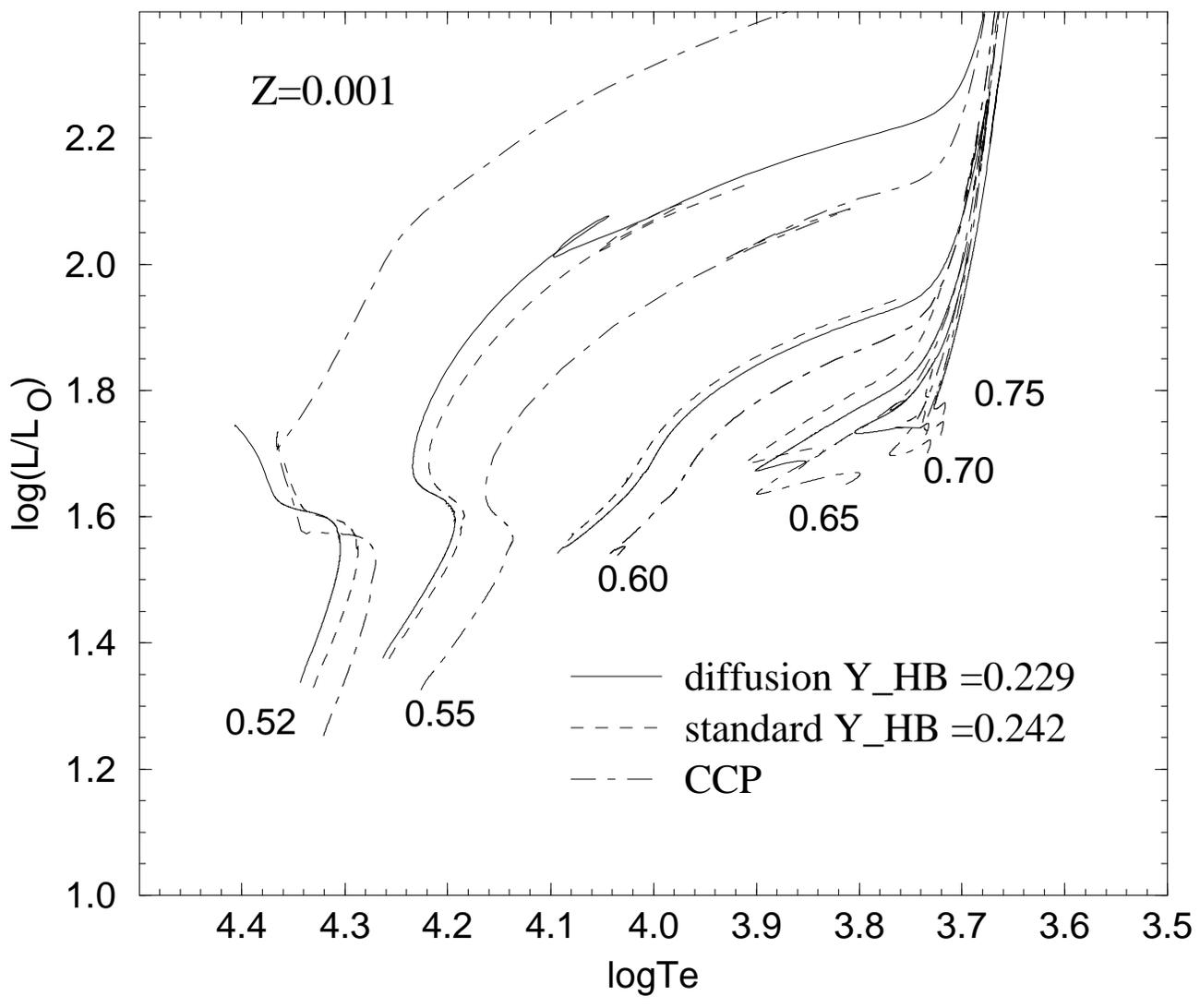
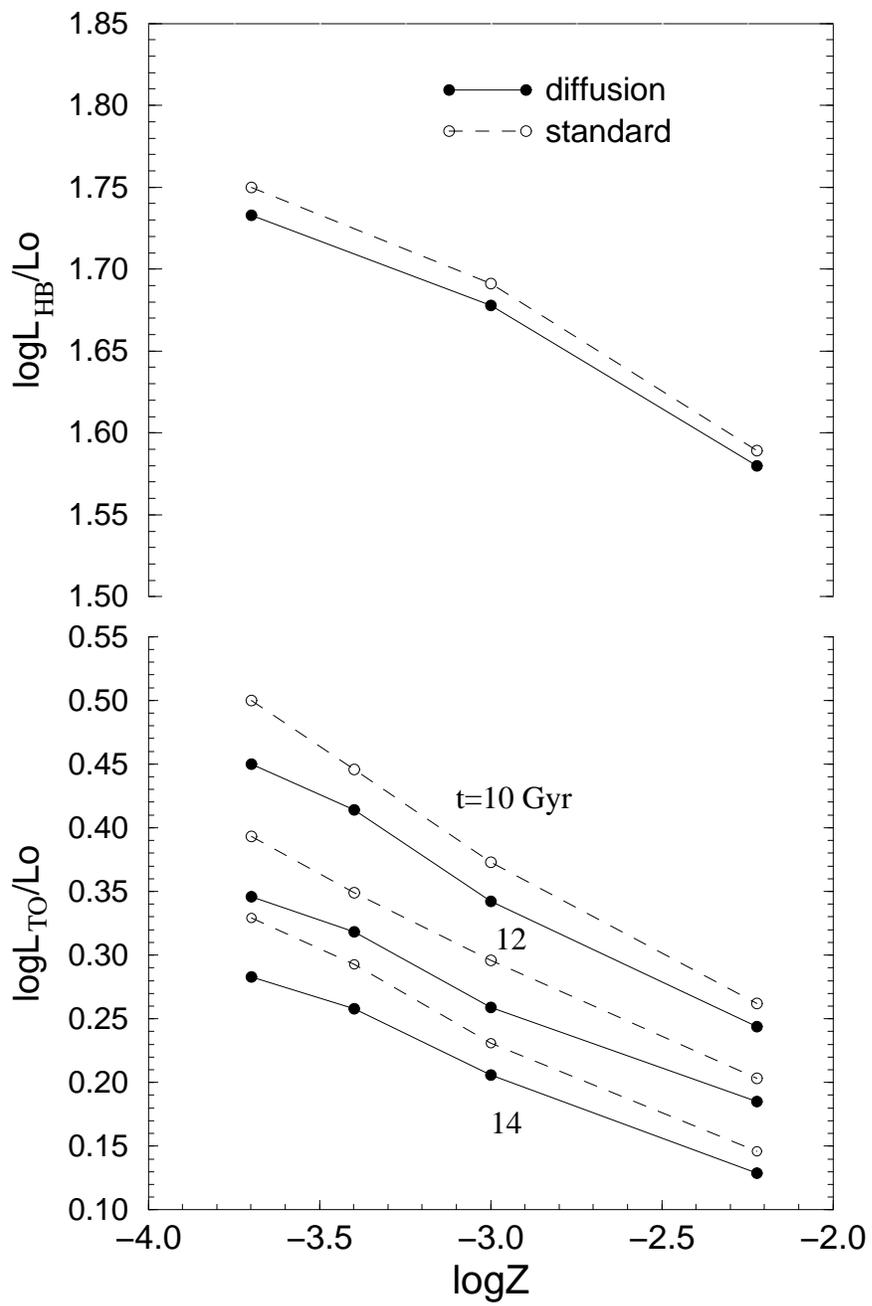
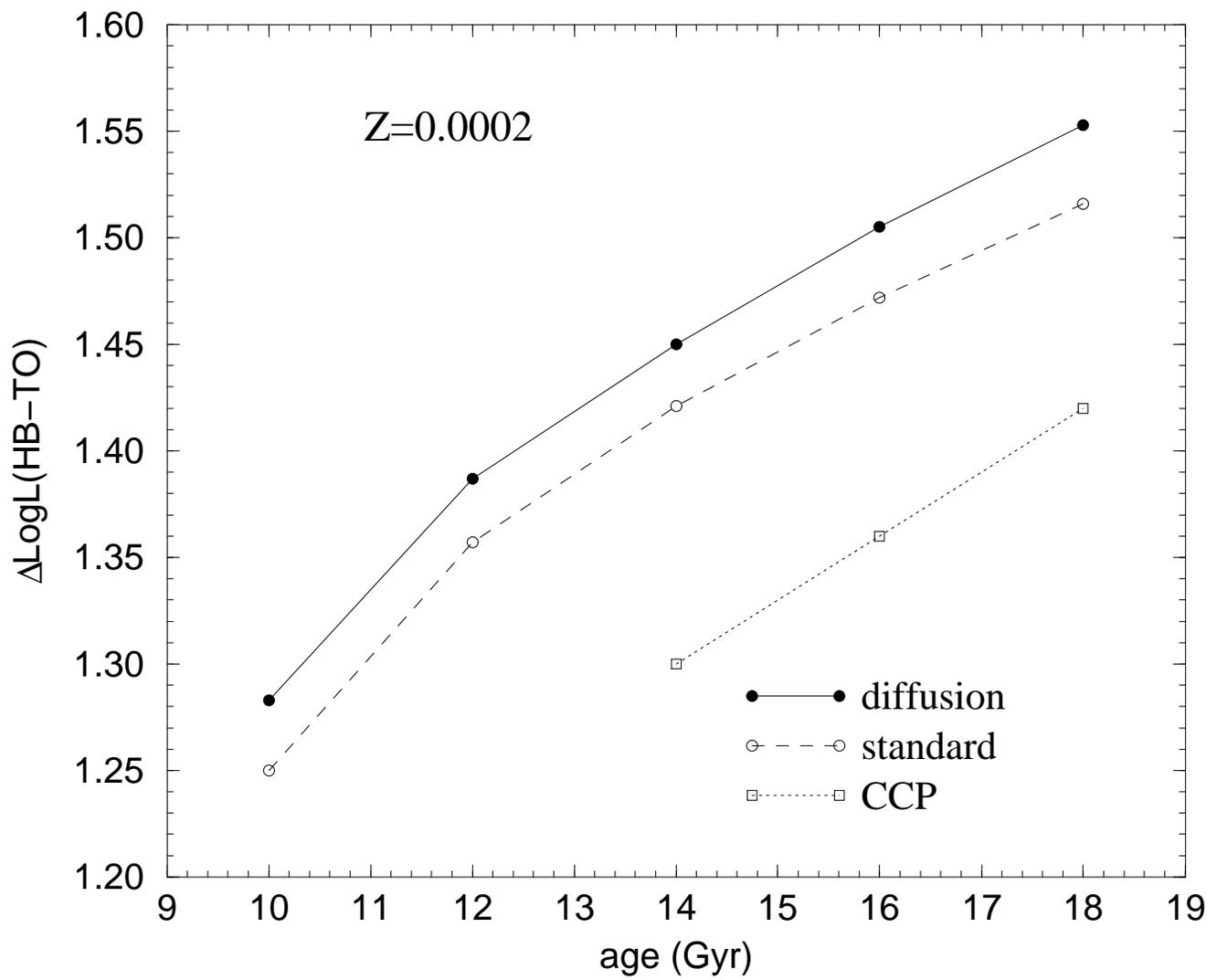
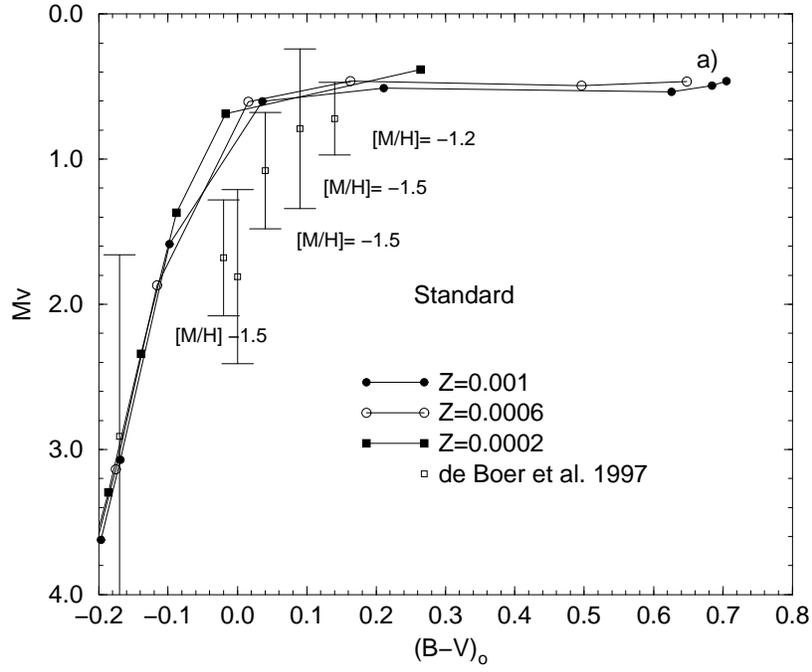
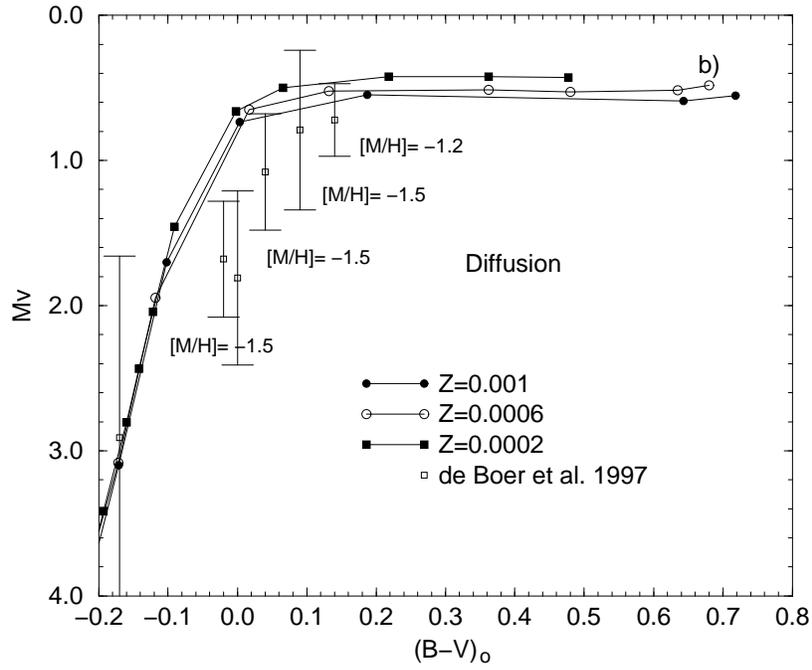
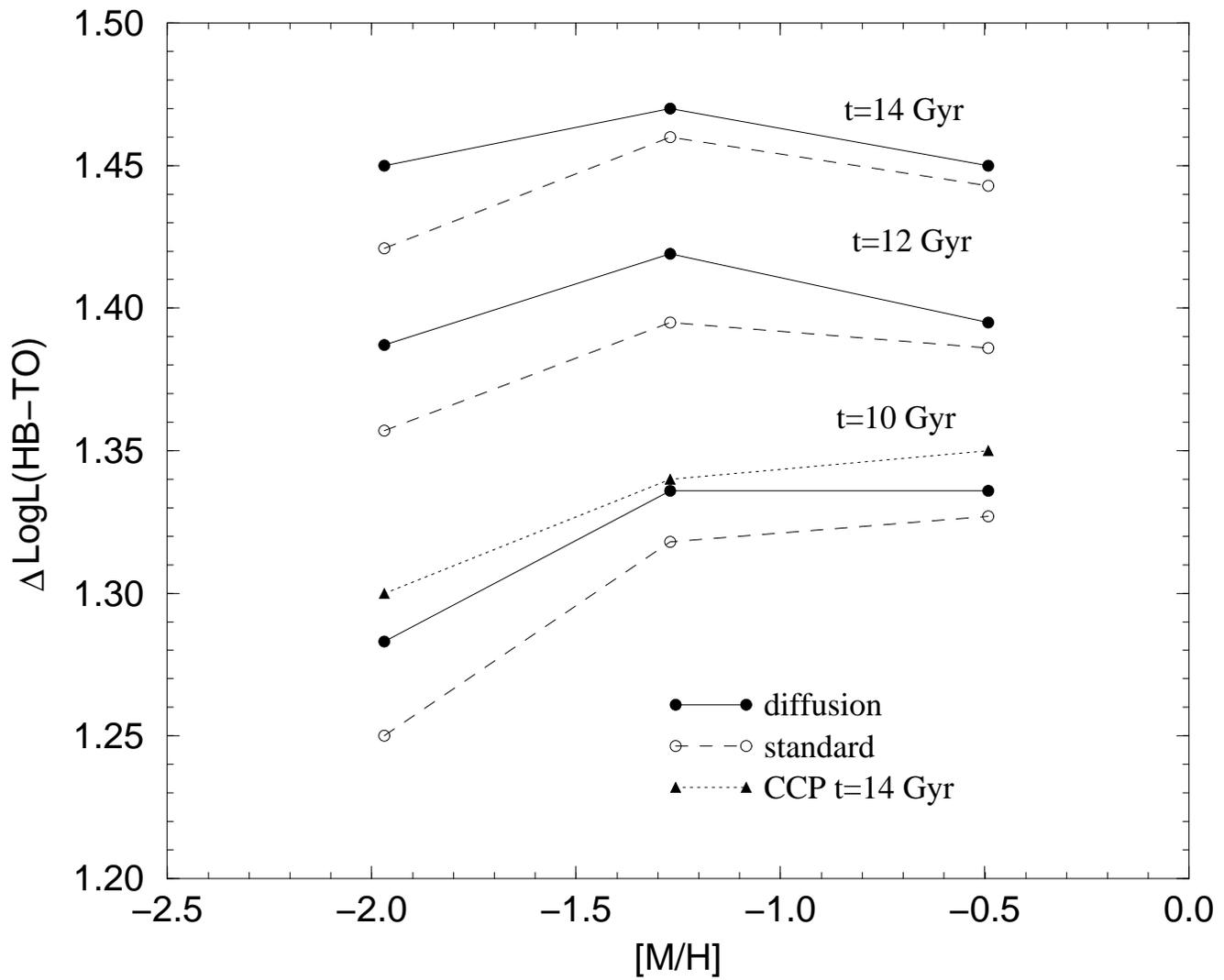
\end{document}


\begin{table}[ht]
\caption[]{Steps in the input physics and related selected evolutionary
quantities for a 0.8 M$_{\odot}$, Y=0.23, Z=0.0001 model, assumed as 
progenitor of the 0.7 M$_{\odot}$ (Y=0.238) HB model at the bottom of
the table. Ages for the 0.8 M$_{\odot}$ model and for the 0.7 M$_{\odot}$
HB model are, respectively, in Gyr and in Myr.}
\begin{center}
\begin{tabular}{l c c c c c c c c}
\hline
\hline
Step & 1      & 2      & 3      & 4    & 5       & 6   &  7    & 8 \\
EOS        & Str88 & Str88 & Str88 & OPAL & OPAL &OPAL &  OPAL & OPAL\\
OPAC       & LAOL & OPAL & OPAL  & OPAL & OPAL   & OPAL & OPAL & OPAL\\
OPAC-CO    & LAOL & LAOL & OPAL  & OPAL & OPAL   & OPAL & OPAL & OPAL\\
$\alpha$-rates & Fow75 & Fow75 & Fow75 & Fow75   & Cau88 & Cau88 & Cau88 & Cau88 \\
NEU & Mun85 & Mun85 & Mun85 & Mun85 & Mun85      & Haft94 & Mun85 & Haft94 \\
\hline\\
LogL$^{TO}$ & 0.384  & 0.382  & 0.382 & 0.411 & -  & - & 0.410 & 0.410\\
t$^{TO}$    & 13.62  & 13.50  & 13.50 & 11.58     & -  & -  & 11.58 & 11.58\\
LogL$^{flash}$ & 3.245 & 3.279 & 3.279 & 3.275 & -  & - & 3.291 & 3.322 \\
t$^{flash}$ & 15.28 & 15.17 & 15.17 & 13.06 & -  &-  & 13.22 & 13.22 \\
M$_c$       & 0.5054 & 0.5098 & 0.5098 & 0.5054 & 0.5054 & 0.5054 & 0.5092 & 0.5152 \\
Y$_{HB}$ & 0.238 & 0.238 & 0.238 & 0.238 & 0.238 & 0.238 & 0.238 & 0.238 \\
\hline\\
LogL$_{ZAHB}$ & 1.617 & 1.635 & 1.635 & 1.656 & 1.663 & 1.663 & - & 1.666\\
LogTe$_{ZAHB}$& 3.976 & 3.991 & 3.990 & 3.983 & 3.974 & 3.974 & - & 4.010 \\
t$_{HB}$      & 96.97 & 93.71 & 87.17 & 87.65 & 80.44 & 80.28 & - & 75.02 \\
\hline
\hline
\end{tabular}
\end{center}
\end{table}

\begin{table}[ht]
\caption[]{Evolutionary properties at the TO for step 8 `best' 
models. The age at the track TO is in Gyr.}
\label{}
\begin{center}
\begin{tabular}{c c c c c}
\hline
\hline\\
Z & M &    t$^{TO}$  &    LogL$^{TO}$ &  LogTe$^{TO}$ \\
\\
\hline\\
0.0001 & 0.8  & 11.6  & 0.410  & 3.826  \\
\hline\\
0.0002 & 0.6  & 33.3  & -0.021  & 3.773 \\
0.0002 & 0.7  & 18.6  & 0.188   & 3.797 \\
0.0002 & 0.8  & 11.2 & 0.378   & 3.824  \\
0.0002 & 0.9  & 7.4  & 0.572  & 3.859 \\
0.0002 & 1.0  & 5.0  & 0.723  & 3.899 \\
\hline\\
0.001 & 0.6 & 35.6 & -0.178 & 3.755 \\
0.001 & 0.7 & 20.0 &  0.060  & 3.777 \\
0.001 & 0.8 & 11.7 & 0.231  & 3.799 \\
0.001 & 0.9 & 7.4  & 0.393  & 3.822 \\
0.001 & 1.0 & 5.0  & 0.577  & 3.852 \\
\hline\\
0.006 & 0.65 & 38.3 & -0.250 & 3.729 \\
0.006 & 0.7 & 29.1  & -0.140  & 3.740 \\
0.006 & 0.8 & 16.8  & 0.017  & 3.759 \\
0.006 & 0.9 & 9.5   & 0.134  & 3.777\\
0.006 & 1.0 & 5.5   & 0.263  & 3.796 \\
\hline
\hline
\end{tabular}
\end{center}
\end{table}

\begin{table}[ht]
\caption[]{Selected TO quantities for H burning isochrones
for the labeled values of metallicity and age. The age at the isochrone TO
is in Gyr.}
\label{}
\begin{center}
\begin{tabular}{c c c c c}
\hline
\hline\\
Z &    t$^{TO}$  &    LogL$^{TO}$  &  LogTe$^{TO}$ & M$^{TO}$ \\
\\
\hline\\
0.0002 &   8 & 0.595   & 3.853      & 0.877 \\
0.0002 &   9 & 0.547   & 3.843      & 0.859 \\
0.0002 &   10	& 0.500   & 3.835      & 0.834 \\
0.0002 &   11 & 0.457   & 3.827      & 0.812 \\
0.0002 &   12 & 0.393   & 3.821      & 0.790 \\
0.0002 &     13 & 0.357   & 3.817      & 0.773 \\
0.0002 &     14 & 0.329   & 3.813      & 0.758 \\
0.0002 &     15 & 0.303   & 3.809      & 0.744 \\
0.0002 &     16 & 0.278   & 3.805      & 0.731 \\
0.0002 &     17 & 0.255   & 3.802      & 0.719 \\
0.0002 &     18 & 0.234   & 3.799      & 0.709 \\
\hline
0.001  &       8&  0.455  &  3.820  &  0.897 \\
0.001 &       9&  0.411  &  3.814  &  0.869 \\
0.001 &      10&  0.373  &  3.809  &  0.845 \\
0.001 &      11&  0.339  &  3.804  &  0.824  \\
0.001 &      12&  0.309  &  3.799  &  0.806  \\
0.001 &      13&  0.256  &  3.796  &  0.786  \\
0.001 &      14&  0.231  &  3.793   &  0.772  \\
0.001 &      15&  0.208  &  3.790  &  0.758  \\
0.001 &      16&  0.187  &  3.787  &  0.746  \\
0.001 &      17&  0.167  &  3.784  &  0.734  \\
0.001 &      18&  0.148  &  3.782  &  0.723 \\
\hline
0.006 &     9.5&  0.281  &  3.781  &  0.932  \\
0.006 &     10 &  0.262  &  3.779  &  0.921  \\
0.006 &     11 &  0.229  &  3.775  &  0.900  \\
0.006 &     12 &  0.203  &  3.772  &  0.881  \\
0.006 &     13 &  0.167  &  3.769  &  0.862  \\
0.006 &     14 &  0.146  &  3.767  &  0.847  \\
0.006 &     15 &  0.114  &  3.764  &  0.831  \\
0.006 &     16 &  0.096  &  3.762  &  0.819  \\
0.006 &     17 &  0.079  &  3.760  &  0.807  \\
0.006 &     18 &  0.059  &  3.757  &  0.796  \\
\hline
\hline
\end{tabular}
\end{center}
\end{table}

\begin{table}[ht]
\caption[]{Selected evolutionary quantities for 0.8 M$_{\odot}$ Red Giant 
models without diffusion. The age at the He flash is in Gyr.}
\label{}
\begin{center}
\begin{tabular}{c c c c c c}
\hline
\hline\\
  Z & M$_c$ & Y$_{HB}$  & t$^{flash}$ & LogL$^{flash}$ & LogL$^{bump}$\\
\\
\hline\\
  0.0001 &  0.515 &0.238  &  13.22  & 3.322  & 2.245  \\      
\hline\\
  0.0002   & 0.511 & 0.239 & 13.22  &  3.349  &  2.168  \\
\hline\\
0.001  & 0.503  & 0.242  & 14.52  & 3.396  & 1.943 \\
\hline\\
0.006  & 0.494 & 0.246  & 20.89  & 3.442  & 1.535 \\
\hline
\hline 
\end{tabular}
\end{center}
\end{table}

\begin{table}[ht]
\caption[]{Selected evolutionary quantities for 
ZAHB models without diffusion. The time spent during the central 
helium burning phase (in Myr) is compared with results from CCP.}
\label{}
\begin{center}
\begin{tabular}{c c c c c}
\hline
\hline\\
\multicolumn{5}{c}{Z=0.0001 Y$_{HB}$=0.238 M$_c$=0.515 M$_{\odot}$}\\
\\
\hline
    M  &    LogL &  LogTe   & t$_{He}$ & t$_{He}^{CCP}$ \\
\hline\\
  0.53 &   1.357 &  4.410 &   93.0 &  -  \\
  0.55 &   1.395 &  4.320 &   88.2 & 113.6 \\
  0.60 &   1.471 &  4.217 &   82.6 & 105.3\\
  0.65 &   1.582 &  4.108 &   78.5 & 100.6 \\
  0.70 &   1.662 &  4.010 &   75.0 & 96.9\\
  0.75 &   1.727 &  3.926 &   74.4 & 94.7 \\
  0.77 &   1.744 &  3.900 &   73.9 &  - \\
  0.80 &   1.769 &  3.860 &    -   & 91.3\\
\hline\\
\multicolumn{5}{c}{Z=0.001 Y$_{HB}$=0.242 M$_c$=0.503 M$_{\odot}$}\\
\\
\hline
    M  &    LogL &  LogTe & t$_{He}$ & t$_{He}^{CCP}$ \\
\hline\\
  0.53 &   1.328 &  4.331 &   98.8 &   - \\
  0.55 &   1.373 &  4.257 &   94.2 & 119.9\\
  0.60 &   1.562 &  4.084 &   86.6 & 111.2\\
  0.65 &   1.688 &  3.855 &   84.0 & 106.8\\
  0.70 &   1.747 &  3.739 &   82.5 & 104.3\\
  0.75 &   1.777 &  3.726 &   81.6 & 101.9\\
  0.80 &   1.794 &  3.722 &    -    &  -   \\
\hline
\hline
\end{tabular}
\end{center}
\end{table}

\begin{table}[ht]
\caption[]{Selected evolutionary quantities for  
models with diffusion at the track TO. The age at the track TO is in Gyr.}
\label{}
\begin{center}
\begin{tabular}{c c c c c}
\hline
\hline\\
\\
Z  & M &    t$^{TO}$  &    LogL$^{TO}$ &  LogTe$^{TO}$ \\
\\
\hline
0.0002 & 0.6  & 30.8 & -0.078  & 3.751  \\
0.0002 & 0.7  & 17.7 &  0.140  & 3.786   \\
0.0002 & 0.8  & 10.9 &  0.330  & 3.814   \\
0.0002 & 0.9  & 7.1  &  0.536  & 3.848   \\
0.0002 & 1.0  & 4.9  &  0.702  & 3.890   \\
\hline
0.001 & 0.7  & 18.7  & 0.026  & 3.770 \\
0.001 & 0.8  & 10.9  & 0.187  & 3.794 \\
0.001 & 0.9  & 6.7   & 0.348  & 3.818 \\
0.001 & 1.0  & 4.8   & 0.551   & 3.847  \\
\hline
0.006 &  0.65 & 35.1 &  -0.255      &   3.721      \\
0.006 &  0.70 & 26.6 &  -0.172      &   3.733      \\
0.006 &  0.80 & 15.4 &  -0.027      &   3.754      \\
0.006 &  0.90 & 8.7  &   0.101      &   3.774       \\
0.006 & 1.0  &  5.1 &  0.245        &   3.794       \\
\hline
\hline
\end{tabular}
\end{center}
\end{table}

\begin{table}[ht]
\caption[]{Selected TO quantities for H burning isochrones 
with diffusion for the labeled assumptions about age and original
chemical composition. The age at the isochrone TO is in Gyr.}
\label{}
\begin{center}
\begin{tabular}{c c c c}
\hline
\hline\\
\multicolumn{4}{c}{Z=0.0002 Y=0.23}\\
\\
\hline
     t$^{TO}$  &    LogL$^{TO}$ &  LogTe$^{TO}$ & M$^{TO}$ \\
\hline
8  &  0.557  & 3.840  &  0.881 \\
9  &  0.501  & 3.831  &  0.852 \\
10 &  0.450  & 3.822  &  0.826  \\
11 &  0.406  & 3.814  &  0.804 \\
12 &  0.346  & 3.809  &  0.783 \\
13 &  0.313  & 3.804  &  0.766 \\
14 &  0.283  & 3.800  & 0.750 \\
15 &  0.255  & 3.796  & 0.736\\
16 &  0.228  & 3.793  & 0.723\\
17 &  0.204  & 3.789  & 0.711\\
18 &  0.180  & 3.786  & 0.700\\
\hline
\multicolumn{4}{c}{Z=0.001 Y=0.23 }\\
\hline
8  &  0.429  & 3.814  &  0.888 \\
9  &  0.383  & 3.807  &  0.861 \\
10 &  0.342  & 3.801  &  0.837  \\
11 &  0.305  & 3.796  &  0.816 \\
12 &  0.259  & 3.791  &  0.795 \\
13 &  0.232  & 3.788  &  0.779 \\
14 &  0.206  & 3.784  & 0.764 \\
15 &  0.183  & 3.781  & 0.750\\
16 &  0.160  & 3.778  & 0.737\\
17 &  0.140  & 3.775  & 0.726\\
18 &  0.120  & 3.773  & 0.715\\
\hline
\multicolumn{4}{c}{Z=0.006 Y=0.23 }\\
\hline
9.5 & 0.265 & 3.776  &  0.926 \\
10 &  0.244  & 3.774  &  0.914  \\
11 &  0.204  & 3.770  &  0.891 \\
12 &  0.185  & 3.767  &  0.874 \\
13 &  0.156  & 3.763  &  0.856 \\
14 &  0.129  & 3.760  &  0.840 \\
15 &  0.090  & 3.757  &  0.823 \\
16 &  0.067  & 3.755  &  0.809 \\
17 &  0.047  & 3.752  &  0.797 \\
18 &  0.031  & 3.750  &  0.785 \\
\hline
\hline
\end{tabular}
\end{center}
\end{table}

\begin{table}[ht]
\caption[]{Selected evolutionary quantities for stellar models
 with element diffusion at the He flash. The age at the He flash 
is in Gyr.}
\label{}
\begin{center}
\begin{tabular}{c c c c c c}
\hline
\hline
  M & M$_c$ & Y$_{HB}$  & t$^{flash}$ & LogL$^{flash}$ & LogL$^{bump}$ \\
\hline\\
\multicolumn{6}{c}{Z=0.0002}\\
\hline\\
  0.7 &  0.5188 &  0.2144 & 20.33  & 3.359 & 2.056  \\
  0.8 &  0.5148 &  0.2261 & 12.77  & 3.352 & 2.139   \\
  0.9 &  0.5113 &  0.2344 & 8.51   & 3.344 & 2.209   \\
  1.0 &  0.5081 &  0.2402 &  5.96  & 3.334 & 2.290   \\
\hline\\  
\multicolumn{6}{c}{Z=0.001}\\
\hline\\
  0.7 &  0.5107 & 0.2178 &  22.24  & 3.410 & 1.825  \\
  0.8 &  0.5068 & 0.2294 &  13.98  & 3.404 & 1.917   \\
  0.9 &  0.5041 & 0.2377 &  9.28   & 3.398 & 1.983   \\
  1.0 &  0.5018 & 0.2434 &  6.46  & 3.393 & 2.043    \\
\hline\\  
\multicolumn{6}{c}{Z=0.006}\\
\hline\\
  0.7 &  0.5034 & 0.2184 &  31.36  & 3.460 & 1.399  \\
  0.8 &  0.5001 & 0.2308 &  19.91  & 3.457 & 1.495   \\
  0.9 &  0.4975 & 0.2392 &  13.30   & 3.457& 1.578   \\
  1.0 &  0.4958 & 0.2446 &  9.20  & 3.530  & 1.657    \\
\hline  
\hline
\end{tabular}
\end{center}
\end{table}

\begin{table}[ht]
\caption[]{Selected evolutionary quantities for 
HB models with diffusion. The time spent during the central He burning phase is in Myr.}
\label{}
\begin{center}
\begin{tabular}{c c c c}
\hline
\hline\\
\multicolumn{4}{c}{Z=0.0002 Y$_{HB}$=0.226 M$_c$=0.515 M$_{\odot}$}\\
\\
\hline
    M  &    LogL &  LogTe & t$_{He}$  \\
\hline\\
  0.53 &   1.352 &  4.395 &   -     \\
  0.54 &   1.369 &  4.352 &    -    \\
  0.55 &   1.386 &  4.318 &   86.6 \\
  0.58 &   1.437 &  4.237 &   -     \\  
  0.60 &   1.478 &  4.192 &   78.1 \\
  0.62 &   1.524 &  4.145 &   78.0 \\  
  0.65 &   1.593 &  4.072 &   76.0 \\
  0.70 &   1.678 &  3.952 &   73.9 \\
  0.72 &   1.703 &  3.909 &     -   \\  
  0.75 &   1.734 &  3.849 &   72.9 \\
  0.77 &   1.751 &  3.813 &     -   \\
  0.80 &   1.773 &  3.772 &   66.6 \\
\hline\\
\multicolumn{4}{c}{Z=0.001 Y$_{HB}$=0.229 M$_c$=0.507 M$_{\odot}$}\\
\\
\hline
    M  &    LogL &  LogTe & t$_{He}$ \\
\hline\\
  0.53 &   1.337 &  4.343 &   95.0\\
  0.55 &   1.376 &  4.263 &   92.8 \\
  0.60 &   1.540 &  4.095 &   84.8 \\
  0.65 &   1.671 &  3.866 &   81.4 \\
  0.70 &   1.730 &  3.735 &   80.3 \\
  0.75 &   1.760 &  3.719 &   78.7  \\
\hline
\hline
\end{tabular}
\end{center}
\end{table}

\begin{table}[ht]
\caption[]{Theoretical estimates of the parameter R. Times are in Myr.}
\label{}
\begin{center}
\begin{tabular}{c c c c c c c}
\hline
\hline
  Z            &  0.0002 &  0.0002 &       0.001 &  0.001 &  0.006 &  0.006 \\
\hline
  Diffusion    &   NO    &  YES    &         NO  &  YES   &   NO   &   YES  \\
\hline\\
  LogL$_{ZAHB}$ & 1.759   &  1.744  &      1.701  &  1.687 &  1.594 &   1.574 \\
  t$_{He}$ & 76.16  &  71.33  &      83.70  &  81.13 &  93.00 &   91.30 \\
  t$_{RG}$ & 73    &  77.24  &      79.52  &  85.22 &  68.3  &   83.20 \\
\hline\\
  R(3.83)      &  1.043   &  0.923  &      1.053  &  0.952 &  1.362 &   1.097 \\
  R(3.83)+0.05 &  1.142   &  1.010  &      1.395  &  1.034 &  1.525 &   1.203 \\
  R(3.83)+0.10 &  1.248   &  1.102  &      1.522  &  1.120 &  1.690 &   1.308 \\
\hline
\hline
\end{tabular}
\end{center}
\end{table}